\documentclass[%
 amsmath,amssymb,
 aps,prl,
 showpacs,
 superscriptaddress,
twocolumn
]{revtex4-2}

\usepackage{cancel}
\usepackage{tikz}
\usepackage{hyperref}
\usepackage{graphicx}
\usepackage{xcolor}
\usepackage{dcolumn}
\usepackage{bm}

\definecolor{lgray}{gray}{0.7}
\newcommand{\red}[1]{\textcolor{black}{{#1}}}

\newcommand{\out}[1]{ }
\begin{document}

\preprint{APS}

\title{
Long-range Order 
and Directional Defect Propagation
in the Nonreciprocal XY Model 
with Vision Cone Interactions
}
\author{Sarah A.M. Loos}
 \email{sl2127@cam.ac.uk}
\affiliation{ 
DAMTP, University of Cambridge, Wilberforce Road, Cambridge CB3 0WA, United Kingdom
}
\author{Sabine H.L. Klapp}
\affiliation{ 
Technische Universit\"at Berlin,
Straße des 17. Juni 135, 10623 Berlin, Germany
}
\author{Thomas Martynec}
\affiliation{ 
Technische Universit\"at Berlin,
Straße des 17. Juni 135, 10623 Berlin, Germany
}

\date{\today}

\begin{abstract}
We study a two-dimensional, nonreciprocal XY model, where each spin interacts only with its nearest neighbours in a certain angle around its current orientation\red{, i.e., its `vision cone'}.
Using energetic arguments and Monte-Carlo simulations we show that a true long-range ordered phase emerges. A necessary ingredient is a configuration-dependent bond dilution entailed by the vision cones. Strikingly, defects propagate in a directional manner, thereby breaking the parity and time-reversal symmetry of the spin dynamics. This is detectable by a non-zero entropy production rate.
\end{abstract}
\maketitle

A growing number of papers 
demonstrates that nonreciprocal (NR) interactions which break the $\textit{actio=reactio}$ principle are the origin of intriguing physical phenomena in nonequilibrium systems~\cite{fruchart2021non,you2020nonreciprocity,saha2020scalar,braverman2021topological,poncet2021soft,loos2020irreversibility,brandenbourger2019non,cavagna2017nonsymmetric,ivlev2015statistical}.
A prominent example are travelling-wave phases in binary fluids, which can be caused by NR coupling between the two fluid components~\cite{you2020nonreciprocity,saha2020scalar}. These time-dependent phases break the 
$\mathcal{PT}$ symmetry of the system, and their emergence has been linked to the existence of underlying exceptional points \cite{fruchart2021non}.  
In solids and soft crystals it was recently shown that NR interactions may introduce odd elasticity~\cite{braverman2021topological,poncet2021soft}. %
A common source of nonreciprocity in biological and artificial systems is perception within a finite \textit{``vision cone''}, which naturally leads to interactions that are NR and orientation-dependent. For example,\red{ which neighbours a 
pedestrian in a crowd \cite{nicolas2021social}, a car driver in a traffic jam \cite{knospe2000towards}, or a bird in a flock \cite{nathan2008v} reacts to}, may depend on its current orientation. 
The few studies on this subject in the area of motile active matter have shown that vision cone interactions can lead to the formation of new self-organized patterns and aggregates~\cite{barberis2016large,durve2018active,costanzo2019milling,lavergne2019group}. 
%

\begin{figure}
	\includegraphics[width=0.9\linewidth]{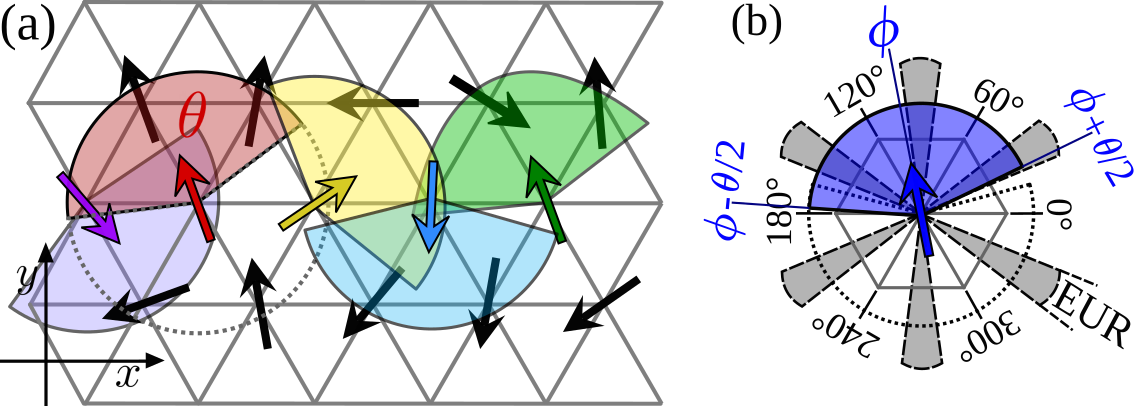}
	\caption{(a) Illustration of the model on a hexagonal lattice. Each spin interacts only with those nearest neighbours lying in its vision cone of size $\theta$. (b) If the spin orientation $\phi$ lies in the EUR (6 grey shaded regions), the total number of coupled neighbours is reduced by one. 
	}
	\label{fig:illustration}
\end{figure}

To gain deeper insights into the physical mechanisms induced by nonreciprocity, we study in this Letter how NR vision cone interactions affect the behavior of many-body systems on a lattice.
Indeed, lattice models have proven invaluable to study fundamental questions of statistical physics, in particular concerning the emergence of phases and phase transitions, 
in addition to having numerous applications in physics, engineering, socioeconomics, and biology.
Here, we implement vision cone interactions into 
the XY model with short-range coupling (Fig. \ref{fig:illustration}), 
which allows us to study the interplay between a continuous rotational dynamics, alignment interactions, and vision cones.
We uncover two intriguing phenomena. 
First, NR interactions can induce a true long-range ordered (LRO) phase. This is in sharp contrast to the standard short-ranged XY model,
in which a LRO phase is forbidden by the Mermin-Wagner theorem.
Remarkably, LRO even arises for vision cones that are almost $360^\circ$.
Second, the vision cone interactions cause defects to propagate in a directional, parity-broken manner. This directional spin dynamics also breaks the time-reversal symmetry, which we measure by the entropy production rate (EPR). Indeed, we have recently shown that nonreciprocity generally causes EPR $>0$ \cite{loos2020irreversibility}. Here, we find that the EPR has a maximum close to the onset of the disordered phase. 
Using a NR version of the classical XY model enables us to rationalize these phenomena by adapting a language and toolbox well-known from equilibrium statistical mechanics, including spin wave 
excitation, energy minimization, and bond percolation. 

%

%

\paragraph{Model.---}
We consider a two-dimensional lattice of spins $S_i \in \mathbb{R}^2$, $i \in \{1,2,....,L^{2}\}$, whose orientations $\phi_i \in [0,360^{\circ})$ can continuously rotate in the lattice plane. All spins are connected to a heat bath at temperature $T$. The total energy of the system is ${E}_\mathrm{tot}=\sum_i {E}_i$ with
\begin{equation}\label{def:H}
{E}_i =  -\sum_{\langle j \rangle} J_{ij}(\phi_i) \cos (\phi_i -\phi_j),
\end{equation}
where the sum runs over all nearest-neighbouring lattice sites of $i$. Different from the standard XY model, the coupling constant $J_{ij}$ explicitly depends on the orientation of spin $S_i$ and the lattice position of spin $S_j$, since only neighbours within the vision cone of size $\theta \in (0,360^{\circ}]$ are coupled, namely
\begin{equation}\label{def:J}
    J_{ij}(\phi_i) = \begin{cases}
    J, & 
    \min \left\{360^\circ\! -|\phi_i - \vartheta_{ij}|, |\phi_i - \vartheta_{ij}| \right\}  \leq \frac{\theta}{2} \\
    0, & \text{else}
    \end{cases},
\end{equation}
where $\vartheta_{ij} \in \{k \,360^\circ/n \}$, $k \in \{1,..,n\}$ denotes the angle of the connecting line between $S_i$ and $S_j$, with $n$ being the number of nearest neighbours per spin. 
\red{As typical representatives for two-dimensional lattice geometries, we consider the hexagonal lattice (where $n=6$) and the square lattice ($n=4$).} In the theoretical considerations, we however do not specify the lattice geometry, so that our results are more general.

We investigate the state-flip dynamics by Monte-Carlo simulations with Glauber transition rates \cite{marro2005nonequilibrium} 
$ w(\phi_{i} \to \phi_i') =  \left[ 1 -  \tanh ( [E_i(\phi_{i})-E_i(\phi_{i}') ]/2T ) \right]/2 $, with the Boltzmann constant set to unity, and apply periodic boundary conditions.
We set $J$ to $1$ (ferromagnetic coupling).
For $\theta = 360^\circ$, the model reduces to the standard XY model.
For $\theta<360^\circ$, the coupling matrix $\textbf{J}$ with elements $J_{ij}$ \eqref{def:J} is generally asymmetric, meaning that the coupling between two spins may be 
NR, $J_{ij}\neq J_{ji}$. In particular,
some \red{bonds are unidirectional ($J_{ij}\neq 0$ but $J_{ji}= 0$)}, and some \red{bonds are missing ($J_{ij}=J_{ji}= 0$)}.
Furthermore, $\textbf{J}$ is {configuration-dependent}.
From the viewpoint of network science, this is a dynamic or temporal network and, due to the presence of NR links, a \textit{directed} graph. However, the fact that $\textbf{J}$ is a function of the orientations $\{\phi_i(t)\}$ crucially
differentiates our model from other spin models on {directed} graphs~\cite{lima2006ising,sanchez2002nonequilibrium,lipowski2015phase}. Here, the spin dynamics and network structure are mutually interrelated.

\out{As global order parameter, we consider the magnetization $m = {L^{-2}} \sum \phi_{i}$. Further, we numerically calculate the susceptibility $\chi = L^{2} T^{-1} \left( \langle m^2 \rangle - \langle m \rangle^2 \right)$, the specific heat 
$C_v = L^{2}T^{-1} \left( \langle E_\mathrm{tot}^2 \rangle - \langle E_\mathrm{tot} \rangle^2 \right)$, 
and to determine $T_{c}$ the fourth-order Binder cumulant (kurtosis) of the magnetization $U_{L} = 1 - \frac{\langle m^4 \rangle}{3 \langle m^2 \rangle^{2}}$ which behaves as $f(\frac{T - T_{c}}{T} L^{1/\nu})$ at criticality (for reciprocal case).}

\begin{figure*}
	\includegraphics[width=\textwidth]{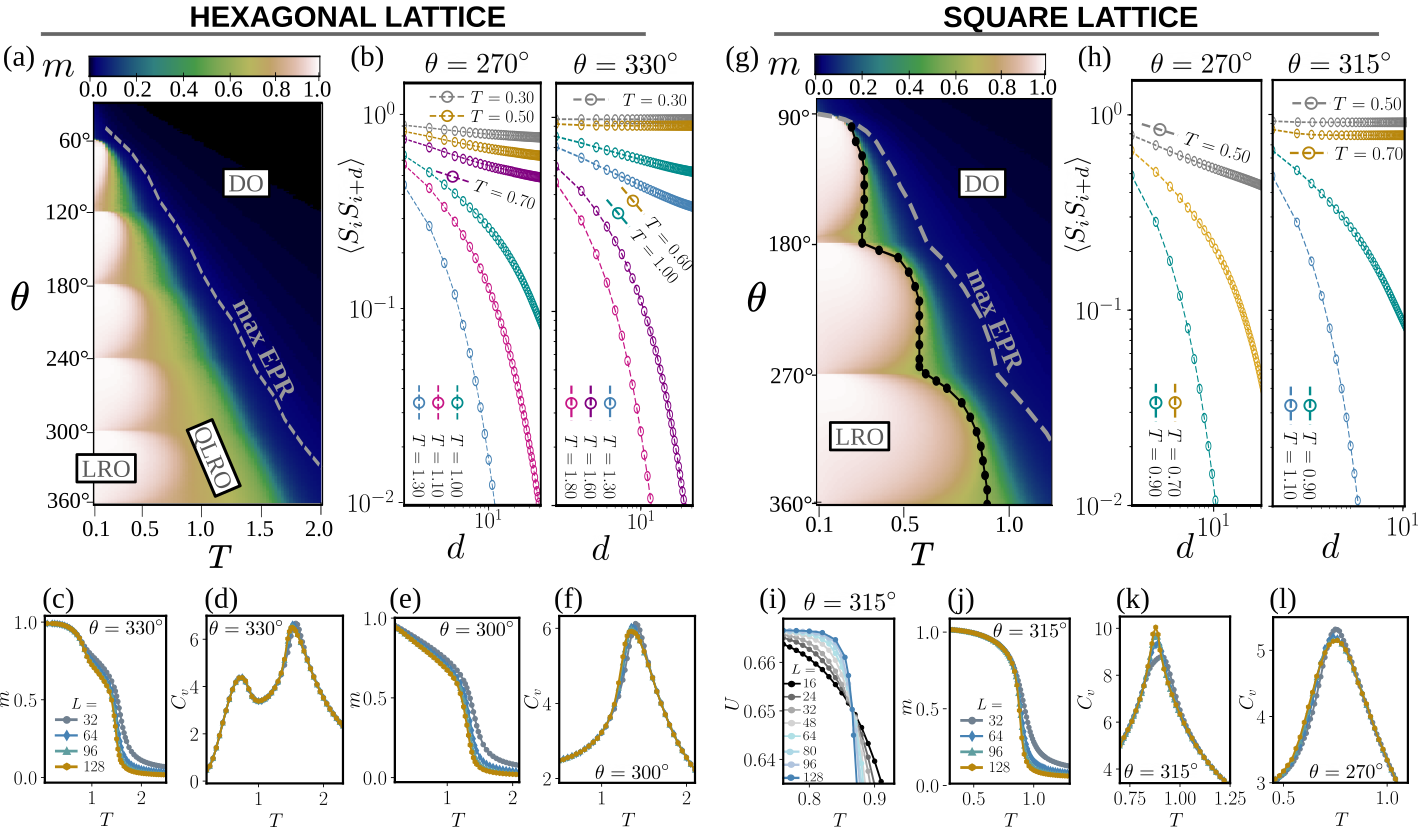}
	\caption{\label{Map} (a-f) \textit{Hexagonal lattice:} (a) Map of magnetization $m =\langle \sqrt{\left( \sum \text{cos} \phi_i \right)^{2} + \left( \sum \text{sin} \phi_i \right)^{2}} / L^{2} \rangle$ as function of $T$ and $\theta$, (b) Decay of spin--spin correlations $\langle S_i S_{i+d} \rangle$ with spin distance $d$ for $\theta = 270^\circ$ and $\theta = 330^\circ$ ($L = 192$). 
	(c,d) $m(T)$ and specific heat $C_v(T)$ for $\theta = 330^\circ$ and various system sizes $L$, (e,f) $m(T)$ and $C_v(T)$ for $\theta = 300^\circ$.
	(g-l) \textit{Square lattice:} (g) Magnetization map, and (h) spin--spin correlations. 
	The black symbols in (g) show the LRO--DO transition temperatures estimated from the crossing points of the Binder cumulants $U=1 - \frac{\langle m^4 \rangle}{3 \langle m^2 \rangle^{2}}$.
	(i,j,k) $U(T)$, $m(T)$, and $C_v(T)$
    for $\theta = 315^\circ$. (l) $C_v(T)$ for $\theta = 270^\circ$.
	}
	\label{fig:phasediagram}
\end{figure*}

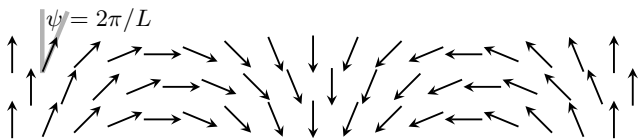
\begin{figure}
\begin{tikzpicture}
 
\def\N{16}%
\def\H{0.43301270189} 
   
\draw[lgray, line width= 0.6mm, rotate around={(-360/\N):(0.5,0.05)}] (0.52,-0.25) -- (0.52,.62); 

\draw[lgray, line width= 0.6mm] (0.4,-0.25) -- (0.4,.6); 

\node[black] (p) at (1.1,0.45) {{  $\psi=2\pi/L$}}; 

\foreach \x in {0,...,\N}
    \draw[-stealth, black, line width= 0.25mm, rotate around={\x*(-360/\N):(0.5*\x,0)}] (0.5*\x,-.25) -- (0.5*\x,.25);

\foreach \x in {0,...,\N}
    \draw[-stealth, black, line width= 0.25mm, rotate around={\x*(-360/\N):(0.25+0.5*\x,-\H)}] (0.25+0.5*\x,-\H-.25) -- (0.25+0.5*\x,-\H+0.25);

\foreach \x in {0,...,\N}
    \draw[-stealth, black, line width= 0.25mm, rotate around={\x*(-360/\N):(0.5*\x,-2*\H)}] (0.5*\x,-2*\H-0.25) -- (0.5*\x,-2*\H+0.25);
\end{tikzpicture}
    \caption{Spin wave of longest wavelength and lowest energy that fully destroys the order in $x$ direction. Here $L=17$. 
    }
    \label{fig:spinwave}
\end{figure}
\paragraph{Phase Behavior.---}
The standard two-dimensional XY model with short-range interactions displays a low temperature quasi-long range ordered (QLRO) phase that crosses at high temperatures by an infinite-order Berezinskii–Kosterlitz–Thouless (BKT) transition to a disordered (DO) phase. Our most striking finding is that for vision cone sizes $\theta < 360^{\circ}$, a LRO ferromagnetic phase at finite $T$ emerges \red{for both lattice geometries}. 
The LRO phase is marked by an average magnetization $m \to 1$ and a large spin--spin correlation $\langle S_i S_{i+d} \rangle$ that spans the entire system (see Fig. \ref{fig:phasediagram}). Averages $\langle .\rangle$ are taken over all spins and multiple realisations. 
\red{For both lattices, the LRO is absent, if $\theta$ is a multiple of
$(360^\circ/n)$, (i.e., a multiple of $60 ^\circ$ for the hexagonal lattice, or of $90 ^\circ$ for the square lattice).}
\red{Besides this, the phase behavior and related transitions differ between both lattices. On the hexagonal lattice,} the system displays a QLRO phase \red{in between the LRO and DO phase}, marked by algebraically decaying \red{$\langle S_iS_{i+d}\rangle$. (In the SM, we also consider a second measure to distinguish the phases.) 
}
%
We detect traces of two corresponding transitions in the form of two peaks in the specific heat
$C_v =  \left( \langle E_\mathrm{tot}^{2} \rangle - 
\langle E_\mathrm{tot} \rangle^{2} \right)/(LT)^{2}$
%
%
[Fig. \ref{fig:phasediagram}(d,f)]. 
The peak heights and positions do not scale with the system size, suggesting that 
both transitions are of infinite-order.
\red{On the square lattice, we do not detect a QLRO phase (unless $\theta$ is a multiple of
$90 ^\circ$, see below). The order--disorder transition on the square lattice} is marked by peaks in $C_v$ that increase with the system size, indicating critical behavior \red{[Fig. \ref{fig:phasediagram}(k)]}. \red{(The Binder cumulants and susceptibility also express behavior typical of second-order phase transitions, see the SM.)} %
\red{For the special cases 
$\theta/(360^\circ/n) \in \mathbb{N}$, we detect on both lattices a QLRO and a DO phase, reminiscent of the phase behavior of the standard XY model.}

To understand this nonequilibrium phase behavior, one feature of the vision cone interactions turns out to be especially important. 
Namely, the \textit{total number} of neighbours $S_j$ of a spin $S_i$ for which $J_{ij}\neq 0$, is generally configuration-dependent. 
In particular, the number is reduced by one, if $\phi_i$ lies in what we call the ``\textit{energetically unfavorable range}'' (EUR).
The EUR nullifies for $\theta$ being multiple of
$(360^\circ/n)$ and otherwise consists of $n$ angular regions [grey areas in Fig. 1(b)] which in total amount to an angle of
$ 
n \, \{ (360^{\circ} /n)-[  \theta \mod (360^{\circ} /n) ]  \}.
$ 
In a LRO configuration, $E_\mathrm{tot}$ is invariant under global rotation of all spins, but only as long as the rotation does not bring $\langle \phi_i\rangle$ in the EUR. Thus, the rotational $U(1)$ symmetry of the system is in general broken. %
Still, the ground state is always infinitely degenerate.
%
If the EUR nullifies, 
%
%
the $U(1)$ symmetry is restored.

Based on energetic considerations, we can now explain the emergence/absence of LRO for different $\theta$.
Consider a fully ordered system. 
In the standard XY model, the energy required to rotate a spin $S_i$ by a small angle $\psi_i$ is $ \mathcal{E}_i \sim \psi_i^2$, \red{which follows from \eqref{def:H}}. In contrast, if $\theta <360^\circ$, the number of bonds can change, yielding $\mathcal{E}_i \sim \psi_i^2 + J$ if the spin enters the EUR due to the rotation, and $\mathcal{E}_i \sim \psi_i^2$ else. 
Rotating all spins by an angle $\psi = 2\pi /L$ w.r.t. their neighbour in one given spatial direction forms the spin wave of longest wavelength that fully destroys the orientational order (Fig.~\ref{fig:spinwave}). 
In cases where the EUR nullifies,
the excitation of such a spin wave takes an energy increase of
$ \mathcal{E}_\mathrm{tot}=\sum_i  \mathcal{E}_i= L^2 4\pi^2 /L^2 = 4\pi^2$, which is finite even in the thermodynamic limit $L\to \infty$. Thus, thermal noise at any arbitrarily small nonzero $T$ can excite this mode and LRO is found only at $T\to 0$. In contrast, if $\theta/(360^\circ/n) \not\in \mathbb{N}$, the energy increase is 
\begin{equation}\label{eq:Etotal}
 \mathcal{E}_\mathrm{tot}= L^2 (4\pi^2 /L^2 + \eta J ) = 4\pi^2 + L^2 \eta J ,
\end{equation} with $\eta \in [0,1)$ denoting the fraction of spins that enter the EUR by the rotation. For large $L$, $\eta$ approaches 
\begin{equation}\label{eq:eta}
\eta \to 1- {[\theta \bmod{ (360^\circ/n)}]}/{(360^\circ/n)},
\end{equation}
which is independent of $L$.
The energetic cost of forming a spin wave thus \textit{diverges} for $L\to \infty$. For large enough systems, this mode cannot be excited by noise and therefore does not destabilize the ground state. This rationalizes the emergence of a LRO phase for $\theta/(360^\circ/n) \not \in \mathbb{N}$.
This argument also explains why the nonreciprocity 
alone does {not} suffice for LRO.

Another observation is the general absence of (Q)LRO phases for narrow cones, $\theta< (360^\circ/n)$. %
This can be explained by a percolation argument. %
The main point is that, as it is well-known from bond-diluted, standard XY models, the bond concentration must be above the bond
percolation threshold $p_\mathrm{c}$ to form QLRO or LRO for finite $T$~\cite{berche2003influence,okabe2005phase,costa2014kosterlitz,kumar2017ordering,kapikranian2012spin,surungan2005kosterlitz}. 
\red{Otherwise, the ground state may comprise disjoint, non-interacting clusters, whose orientations are independent from each other, preventing global ordering.}
In the NR XY model with vision cone $\theta<(360^\circ/n)$, 
the \red{total} fraction of bonds \red{(i.e., edges $\{i,j\}$} with $J_{ij}\neq 0$ and/or $J_{ji}\neq 0$) lies within $p\in [0,1/3]$ for the hexagonal, and $p\in [0,1/2]$ for the square lattice (\red{more details are provided in the SM}). In both cases, $p$ does not exceed the respective bond
peculation threshold, which is 
$p_\mathrm{c}= 2 \sin (\pi/18)
\approx 0.347$ for the hexagonal and
$p_\mathrm{c}= 1/2$ for the square lattice ~\cite{sykes1964exact}. 
Remarkably, it is not the concentration of bidirectional bonds $p_\mathrm{bi}$ \red{(i.e., bonds where $J_{ij}\neq 0$ \textit{and} $J_{ji}\neq 0$)} that matters. 
(Indeed, $p_\mathrm{bi}> p_\mathrm{c}$ only holds for 
$\theta \geq 240^\circ$ on the hexagonal and
$\theta \geq 270^\circ$ on the square lattice.)
Rather, a total bond concentration $p$ of bi- {and} unidirectional bonds above $p_c$ suffices. \red{This is in sharp contrast} to bond diluted spin models on random, directed graphs \cite{lipowski2015phase}. This is because, in our model, the existence/absence of each bond is not random, but \red{determined by} the spin orientations. In the LRO phase, the system self-organizes such that exactly those bonds are present that are needed for percolation.
%
%
%

Next, we consider the $\theta$ dependence of the transition temperatures. %
First, as $\theta$ decreases, the transition to DO is overall shifted towards lower temperatures (Fig.~\ref{fig:phasediagram}). This suits to the fact that $p$ decreases with $\theta$. Indeed, for the reciprocal bond-diluted XY model, the BKT-transition is known to decrease with $p$~\cite{surungan2005kosterlitz}. 
Next, 
the low-temperature transition from LRO to QLRO or DO is found to depend non-monotonically on $\theta$. 
Although this is a nonequilibrium system, the trend can be rationalized by considering the balance of energy and ``configuration entropy''. %
First, the ground state energy decreases with decreasing $\theta$. Hence, the energetic benefit of forming a LRO phase decreases, and the transition temperature is reduced.
Second, for $\theta$ values between two multiples of $(360^\circ/n)$, the number of bonds and thus the energy of the fully ordered ground state is independent of $\theta$. However, the EUR increases linearly with decreasing $\theta$. 
Assuming that the entropy is a logarithmic function of the number of accessible microstates, i.e., spin orientations, in the ground state, this implies that the entropic freedom due to spin rotation without entering the EUR is decreasing logarithmically. Correspondingly, the transition temperature is lowest close to the $\theta$ values, where the EUR is largest (see the SM for illustrations).

\begin{figure}
	\includegraphics[width=0.99\linewidth]{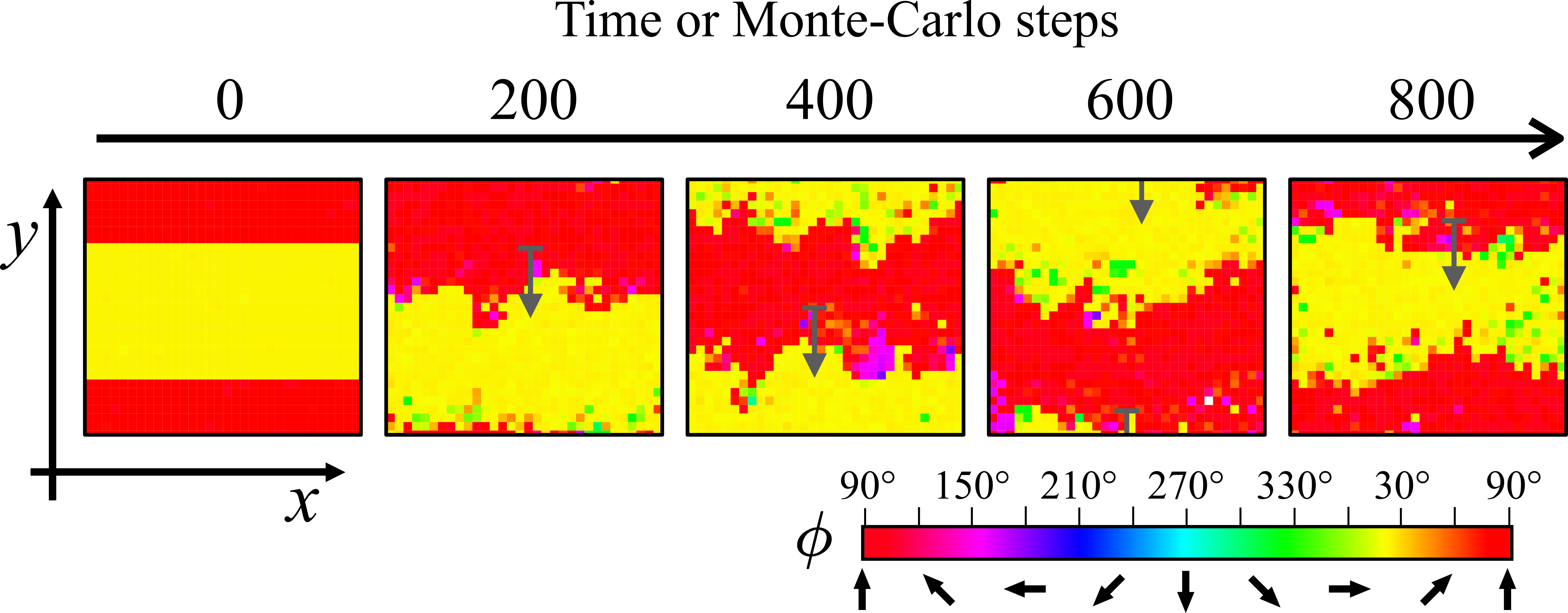}
    \caption{Illustration of defect propagation. 
    Left panel: initial configuration with spin orientations $\phi=30^\circ$ in the yellow middle band, and $\phi=90^\circ$ else. The two interfaces (defect lines) propagate downwards as shown in the consecutive panels. Here, $L = 32$, $T = 0.1$, $\theta = 65^\circ$, hexagonal lattice.}
    \label{fig:PT-SpinwavePropagation}
\end{figure}

\paragraph{Time-reversal symmetry breaking.--} It has recently been established that NR interactions can %
induce stable time-dependent phases, such as travelling waves~\cite{fruchart2021non,you2020nonreciprocity,saha2020scalar}. These phases emerge in systems that comprise two different species with a ``run-n-catch" or ``prey-predator" relation, i.e., one species is attracted to/aligning with the other, while the reverse interaction is repulsive/anti-aligning~\cite{fruchart2021non,you2020nonreciprocity,saha2020scalar}. In contrast, our model comprises only a \textit{single} species of constituents who all interact with each other in an identical manner. %
We do not observe time-dependent phases, and all emergent phases have an equilibrium counterpart.
However, we find that the spin dynamics in all three phases clearly reveals the far-from-equilibrium character of the system. In particular, we observe long-lived (yet transient) directional propagation of local heterogeneous spin structures, e.g., defects. 
To visualise the directional propagation, we prepare the system as follows. We start from a fully ordered configuration in the LRO phase and at time zero introduce two defect lines, see Fig. \ref{fig:PT-SpinwavePropagation}. In marked contrast to an equilibrium system, where the defects would diffuse in all spatial directions,
the defect lines propagate in a preferred spatial direction. The direction of propagation is dictated by the \red{spin} orientations. The underlying reason is that the defect is ``visible'' only to the spins facing it, meaning that the information travels only in certain spatial directions. Despite the spatial symmetry of the initial bands (here, the parity symmetry, $y \to -y$), the response of the system is strongly asymmetric, and breaks the parity. 
Due to the predetermined direction of propagation, the initiated spin dynamics moreover clearly breaks the \textit{time-reversal symmetry}. Indeed, if the video was played backwards the defect lines would move in the ``wrong'' direction. The initiated ``travelling wave'' survives a considerable period of time (for the parameters in Fig.~\ref{fig:PT-SpinwavePropagation}, the defects travel $\sim 10$ times over $L$ before dissolving). 
Eventually, the system collectively reorganizes into a LRO state which lacks global currents.

\begin{figure}
	\includegraphics[width=0.99\linewidth]{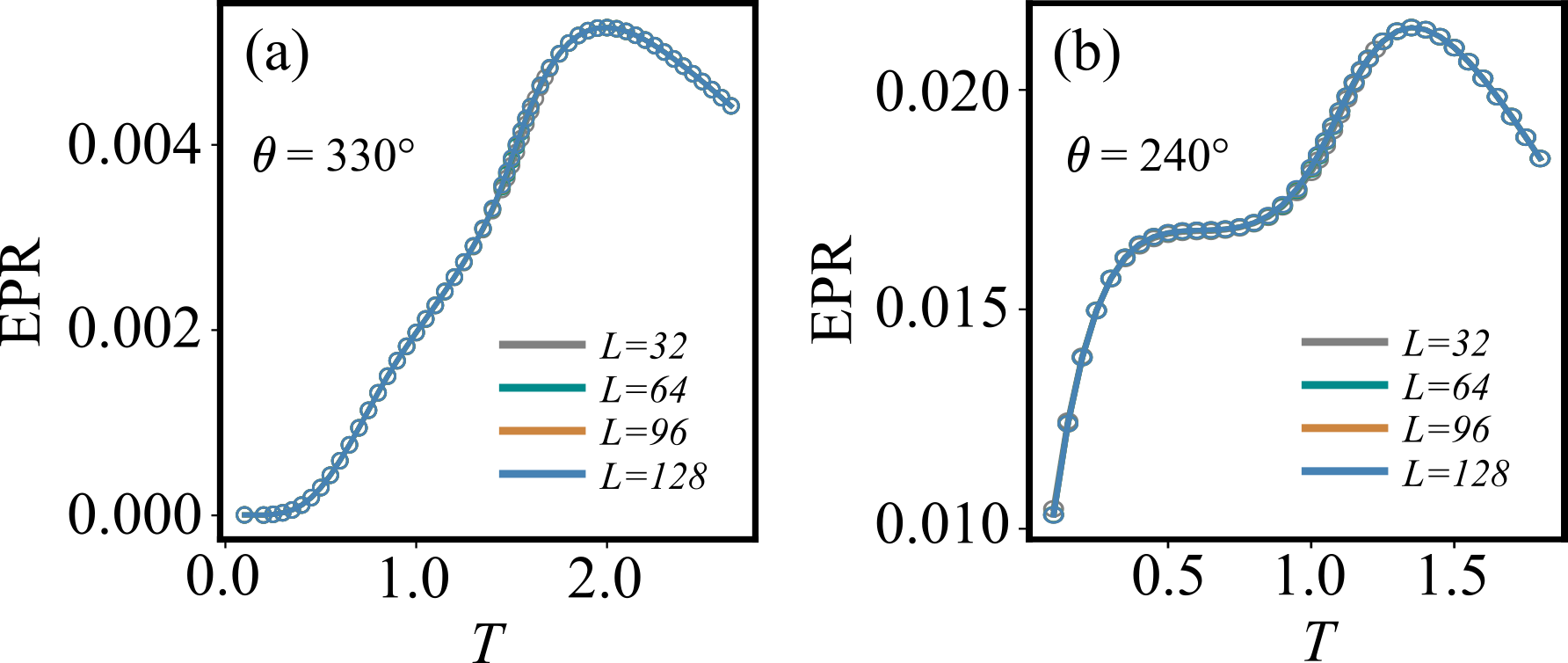}
	\caption{\label{fig:EP} Entropy production rate \eqref{def:EPR} versus $T$ on the hexagonal lattice with (a) $\theta = 330^\circ$, and (b) $\theta = 240^\circ$.
	$L=192$.
	}
\end{figure}
%
 
However, the mechanism described above leads to directional, fluctuating spin dynamics in all three phases. The resulting time-reversal symmetry breaking can be quantified by the mean entropy production rate (EPR) per spin, which we numerically evaluate using a formula from stochastic thermodynamics~\cite{tome2012entropy,schnakenberg1976network,noa2019entropy,martynec2020entropy}
\begin{equation}\label{def:EPR}
    \text{EPR}= 
    \left\langle 
    \text{ln} \frac{w[\phi_i(t) \to \phi_{i}(t+\mathrm{d}t)]}{w[\phi_{i}(t+\mathrm{d}t) \to \phi_{i}(t)]} \right\rangle/\mathrm{d}t.
\end{equation}
Note that \eqref{def:EPR} generally measures the medium EPR (i.e., the dissipated heat over $T$), which in steady states coincides with the {total} EPR.
We find a strictly positive EPR in all phases (Fig. \ref{fig:EP}). This is true for all $\theta<360^\circ$, even for those $\theta$ values, where the EUR vanishes and the number of bonds is constant [Fig. \ref{fig:EP}(b)]. Interestingly, the EPR has a pronounced maximum, which always lies within and close to the onset of the DO phase. The location of this maximum for all $\theta$ values is indicated in Fig. \ref{fig:phasediagram}(a) by the grey dashed line. 
The maximum of the EPR can be explained by the competition of two effects. On the one hand, in view of the directional propagation of defects, which we identified as a source of time-reversal symmetry breaking, it makes sense that the EPR generally increases with $T$, simply because the defect density increases. This also explains why the increase of EPR is particularly steep around the transitions. On the other hand, the defect density eventually saturates in the DO phase, and the orientation of each spin decorrelates with its local environment.
Upon a further raise of $T$, 
the noise then only dominates more and more over the increasingly irrelevant alignment interactions and thus overshadows the propagation mechanism. The result is an ultimately purely random motion -- which is time-symmetric.
Lastly, we note that the slope of EPR has a similar behavior as $C_v$, which was also observed in \cite{martynec2020entropy,tome2012entropy,noa2019entropy}, and that the EPR fluctuations differ qualitatively in the (Q)LRO phases from the DO phase (see the SM).

\paragraph{Discussion.---}
We provided numerical evidence and analytical reasoning 
demonstrating that NR vision cone interactions can lead to the emergence of a LRO phase in a two-dimensional model of continuous spins with short-range coupling. 
The nonreciprocity alone is not sufficient,
but the orientation-dependent bond dilution is crucial. \red{Furthermore, the nature of the phase transitions is different on the square and hexagonal lattice.}
Interestingly, in a coarse-grained model with NR coupling and spin inertia, the theoretical existence of LRO was recently shown on a hydrodynamic level~\cite{dadhichi2020nonmutual}, suggesting an alternative mechanism to introduce LRO in NR systems.
We further found that local heterogeneous spin structures travel with a preferred direction. 
Somewhat analogously, NR interactions in (soft) crystals can turn topological defects in the crystalline structure into motile objects~\cite{braverman2021topological,poncet2021soft}. 
Since the here described mechanism is specific to the vision cone interactions, a further investigation of the relation between both phenomena could yield valuable insights.
We have also studied the time-reversal symmetry breaking of the spin dynamics by the EPR, and found a maximum close to the transition to DO. To obtain a more profound understanding it could be worthwhile to investigate from an entropic and dynamical perspective, the formation, annihilation and motion of vortex--antivortex pairs in the presence of NR coupling.
Another interesting research perspective concerns the relation to motile active matter. 
{The directional propagation of defects and information described here could have a drastic impact on flock cohesion and collective turns \cite{cavagna2017nonsymmetric,Attanasi2014information}.}

\begin{acknowledgments}
This work was funded by the Deutsche Forschungsgemeinschaft (DFG, German Research Foundation) -- through the projects 498288081;  and 163436311-SFB 910. We thank Benjamin Walter, Asja Jeli\'c, Gianmaria Falasco, Jaron Kent-Dobias, Cesare Nardini, Tal Agranov, and Michel Fruchart for valuable comments. 
\end{acknowledgments}

\bibliography{bibliography.bib} 

\end{document}